\documentstyle[sprocl]{article}

\input{psfig}

\def\Journal#1#2#3#4{{#1} {\bf #2}, #3 (#4)}

\def\NPB{{\em Nucl. Phys.} B}
\def\PLB{{\em Phys. Lett.}  B}

\def\PRD{{\em Phys. Rev.} D}

\def\be{\begin{equation}}
\def\ee{\end{equation}}
\def\bea{\begin{eqnarray}}
\def\eea{\end{eqnarray}}

\begin{document}
\begin{flushright}
 July 23  1998\\
JLAB-THY-98-28\\
\end{flushright}

\medskip
\title{FACTORIZATION AND EFFECTIVE ACTION FOR HIGH-ENERGY SCATTERING IN QCD}

\author{IAN BALITSKY}
\address{ Physics Department, Old Dominion University, Norfolk 
VA 23529 \\and \\
Theory Group, Jefferson Lab, Newport News VA 23606\\
e-mail: balitsky@jlab.org} 

\maketitle
\centerline{}

{\it Talk given at ``Continious Advances in QCD", Minnesota, April 16-19,
1998}

\bigskip
\abstracts{ I demonstrate that the amplitude of the 
high-energy scattering can be factorized in a convolution of the contributions 
due to fast and slow fields. The fast and slow fields interact
by means of Wilson-line operators -- infinite gauge factors ordered
along the straight line. The resulting factorization formula gives
a starting point for a new approach to the effective action for 
high-energy scattering.}

\section{Introduction}
The starting point of almost every perturbative QCD calculation is a 
factorization formula of some sort. A classical example is the
factorization of the structure functions of deep inelastic scattering 
into coefficient functions and parton densities. The form of 
factorization is dictated by process kinematics 
(for a review, see\cite{fak}).  
In case of deep inelastic 
scattering, there are two different
scales of transverse momentum and it is therefore natural to 
factorize the amplitude in the product of contributions of 
hard and soft parts coming from the regions of small and large transverse 
momenta, respectively. On the contrary, in the case of high-energy 
(Regge-type) processes, all the transverse momenta are of the same order of 
magnitude,  but colliding particles strongly differ in rapidity. 
Consequently, it is natural to look for factorization in the
rapidity space.

The basic result of the paper is that the 
high-energy scattering amplitude can be factorized in a convolution of 
contributions due to ``fast" and ``slow" fields. To be precise, we 
choose a certain rapidity $\eta_0$   to be a ``rapidity divide" 
and we call
fields with $\eta>\eta_0$ fast and fields with $\eta<\eta_0$ slow 
where $\eta_0$ lies in the region between spectator 
rapidity and target rapidity. (The interpretation of this fields as
fast and slow is literally true only
in the rest frame of the target but we will use this 
terminology for any frame).

Our starting point is the operator expansion for high-energy scattering 
\cite{ing}
where the explicit integration over fast fields gives the coefficient
functions for the Wilson-line operators representing the 
integrals over slow fields. For a 2$\Rightarrow$2 particle
scattering in Regge limit $s\gg m^2$ 
(where $m$ is
a common mass scale for all other momenta in the problem 
$t\sim p_A^2 
 \sim (p'_A)^2\sim p_B^2\sim (p'_B)^2\sim m^2$)
  we have:
\begin{eqnarray}
\lefteqn{A(p_A,p_B\Rightarrow p'_A,p'_B)=}
\label{fla1}\\
&\sum\int d^2x_1...d^2x_nC^{i_1...i_n}(x_1,...x_n)
\langle p_B|{\rm Tr}\{U_{i_1}(x_1)...U_{i_n}(x_n)\}|p'_B\rangle
\nonumber
\end{eqnarray}
(As usual, $s=(p_A+p_B)^2$ and $t=(p_A-p'_A)^2$).
Here  $x_i~(i=1,2)$ are the transverse coordinates 
(orthogonal to both $p_{A}$ and $p_{B}$) and 
$U_i(x)=U^{\dagger}(x){i\over g}{\partial\over\partial x_i}U(x)$ where 
the Wilson-line operator $U(x)$ is the 
gauge link ordered along the infinite
straight line corresponding to the ``rapidity divide'' $\eta_0$. Both 
coefficient functions and matrix elements in Eq. (\ref{fla1}) depend 
 on the $\eta_0$ but 
this dependence is canceled in the physical amplitude just as the scale 
$\mu$ (separating coefficient functions and matrix elements) disappears 
from the final results for structure functions in case of usual factorization.
Typically, we have the factors $\sim (g^2\ln s/m^2-\eta_0)$ coming from
the ``fast" integral and the factors $\sim g^2\eta_0$ coming from
the ``slow" integral so they combine in a usual log factor
$g^2\ln s/m^2$. In the leading log approximation these factors
sum up into the BFKL pomeron\cite{bl},\cite{fkl} (for a review 
see ref. \cite{lobzor}). 
Note, however, that unlike usual factorization,
the expansion (\ref{fla1}) does not have 
the additional meaning of perturbative $vs$ nonperturbative separation 
-- both the coefficient
functions and the matrix elements have perturbative and 
non-perturbative parts. This happens due to the fact that  the 
coupling constant in a
scattering processis is determined by 
the scale of transverse momenta. When we perform 
the usual factorization 
in hard ($k_{\perp}>\mu$) and soft ($k_{\perp}<\mu$) momenta, 
we calculate the 
coefficient functions perturbatively (because 
$\alpha_s(k_{\perp}>\mu)$ is small) whereas
the matrix elements are non-perturbative. Conversely, when we factorize 
the amplitude in rapidity, both fast and slow parts have 
contributions coming from the regions of large and small 
$k_{\perp}$. In this 
sense, coefficient functions and matrix elements enter the expansion 
(\ref{fla1}) on equal footing. We could have integrated first over 
slow fields (having the rapidities close to that of $p_B$)
and the expansion would have the form:
\begin{eqnarray}
&A(s,t)=\sum\int d^2x_1...d^2x_nD^{i_1...i_n}(x_1,...x_n)
\langle p_A|{\rm Tr}\{U_{i_1}(x_1)...U_{i_n}(x_n)\}|p'_A\rangle
\label{fla2}
\end{eqnarray}
In this case, the coefficient functions $D$ are the results of integration 
over slow fields ant the matrix elements of the $U$ operators contain only the
large rapidities $\eta >\eta_0$. The symmetry between 
Eqs. (1) and (2)
calls for a factorization formula which would have this symmetry between 
slow and fast fields in explicit form. 

Our goal is to demonstrate that one can combine the operator expansions 
(\ref{fla1}) and (\ref{fla2}) in the following way:
\begin{eqnarray}
\lefteqn{A(s,t)=\sum{i^n\over n!}\int d^2x_1...d^2x_n}\label{fla3}\\
&\langle p_A|U^{a_1i_1}(x_1)...U^{a_ni_n}(x_n)|p'_A\rangle 
\langle p_B|U^{a_1}_{i_1}(x_1)...U^{a_n}_{i_n}(x_n)|p'_B\rangle
\nonumber
\end{eqnarray}
where $U^a_i\equiv {\mathop{\rm Tr}}(\lambda^aU_i)$ ($\lambda^a$ 
are the Gell-Mann matrices). It is possible 
to rewrite this factorization 
formula in a more visual form if we agree that operators 
$U$ act only on states 
$B$ and $B'$ and introduce the notation $V_i$ for the same operator as 
$U_i$ only acting on the $A$ and $A'$ states:
\begin{eqnarray}
&A(s,t)=\langle p_A|\langle p_B|
\exp\left(i\!\int\! d^2xV^{ai}(x)
U^a_i(x)\right)|p'_A\rangle|p'_B\rangle
\label{fla4}
\end{eqnarray}
\begin{figure}[h]
\hspace{4cm}\psfig{figure=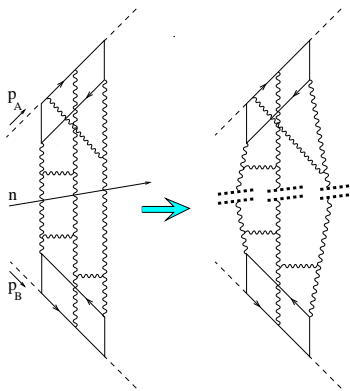,height=2in}
\caption{Structure of the factorization formula. Dashed, solid, and
wavy lines denote photons, quarks, and gluons, respectively. Wilson-line
operators are denoted by dotted lines and the vector $n$ gives the direction of
the ``rapidity divide" between fast and slow fields\label{fig:radish}}
\end{figure}
In a sense, this formula amounts
to writing the coefficient functions in 
Eq. (\ref{fla1}) (or Eq. (\ref{fla2})) 
as matrix elements of 
Wilson-line operators. (Such an idea was first discussed 
in ref. \cite{collell}).
Eq. (\ref{fla4}) illustrated in Fig.1 is our main result 
and the rest of the paper
is devoted to the derivation of this formula and the 
discussion of its possible applications.

\section{Operator expansion for high-energy scattering}
Let us now briefly remind how to obtain the operator expansion
(\ref{fla1}). For simplicity, 
consider the classical example of high-energy scattering of 
virtual photons with virtualities $\sim -~m^2$. 
\begin{equation}A(s,t)=-i{\mbox{$\langle 0|$}} 
T\{j(p_A)j(p'_A)j(p_B)j(p'_B)\}{\mbox{$|0\rangle $}}.
\label{fla5}
\end{equation}
where $j(p)$ is the Fourier transform of 
electromagnetic current $j_{\mu}(x)$ 
multiplied by some suitable polarization $e^{\mu}(p)$.
At high energies it is convenient to use the Sudakov decomposition:  
\begin{equation}
p^{\mu}~=~\alpha_pp_1^{\mu}+\beta_pp_2^{\mu}+p_{\perp}^{\mu}
\label{suda}
\end{equation}
where
$p_1^{\mu}$ and
$p_2^{\mu}$ are the
light-like vectors close to $p_A$ and 
$p_B$, respectively 
($p_A^{\mu}= p_1^{\mu}-p_2^{\mu}p_A^2/s,~
p_B^{\mu}= p_2^{\mu}-p_1^{\mu}p_B^2/s$).
We want to integrate over the fields with 
$\alpha>\sigma$ where $\sigma$ is
defined in such a way that the corresponding 
rapidity is $\eta_0$. (In explicit form  
$\eta_0=\ln{\sigma\over\tilde{\sigma}}$
where $\tilde{\sigma}\equiv {m^2\over s\sigma}$). 
The result of the integration
will be given by Green functions of the fast particles in 
slow ``external" fields\cite{ing} (see also ref.\cite{larry1}).
 Since the fast particle moves along a 
straight-line classical trajectory, 
the propagator is proportional to 
the straight-line ordered gauge factor $U$ \cite{nacht}. For example, when 
$x_{+}>0,~y_{+}<0$ it has the form\cite{ing}:
\begin{eqnarray}
G(x,y)=i\int \! dz\delta
(z_{\ast})
\frac {(\not\!\! x-\not\!\! z)\not\!\!p_2}{2\pi ^{2}(x-z)^{4}}
U(z_{\perp})\frac {\not\!\! z-\not\!\! y}{2\pi
^{2}(z-y)^{4}}
\label{fla6}
\end{eqnarray}
We use the notations 
$z_{\bullet}\equiv z_{\mu}p_1^{\mu}$ and $z_{\ast}\equiv z_{\mu}p_
2^{\mu}$ which 
are essentially identical to the light-front coordinates $z_+=z_*/\sqrt{s},
~z_-=z_{\bullet}/\sqrt{s}$. 
The Wilson-line operator $U$ is defined as
\begin{equation}
U(x_{\perp})=[\infty p_1+x_{\perp}, -\infty p_1+x_{\perp}]
\label{fla7}
\end{equation}
where $[x,y]$ is the straight-line ordered gauge link 
suspended between the points $x$ and $y$:
\begin{eqnarray}
&[x,y]{\stackrel{\rm def}\equiv}
 P\exp\left(ig\int_0^1du (x-y)^{\mu}A_{\mu}(ux+(1-u)y)\right)
\label{fla8}
\end{eqnarray}

The origin of Eq. (\ref{fla6}) is more clear in the rest 
frame of the ``A" photon (see Fig.2). 
\begin{figure}[h]
\hspace{4cm}\psfig{figure=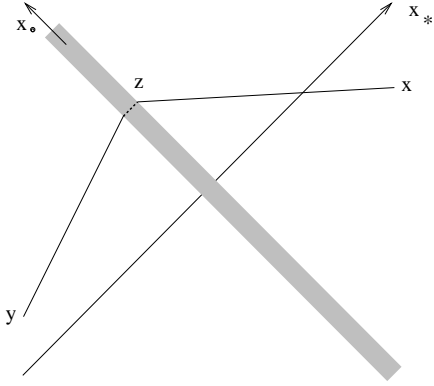,height=2in}
\caption{Quark propagator in the shock-wave background}
\end{figure}
fields are approaching this quark at high speed. Due to the Lorentz
contraction, these fields are squeezed in a shock wave located at $z_{*}=0$.
Therefore,  the propagator (\ref{fla6}) of the quark in this
shock-wave background is a product of three factors which reflect 
(i) free propagation 
from $x$ to the shock wave (ii) instantaneous interaction with the shock 
wave which
is described by the operator $U(z_{\perp})$, and 
(iii) free propagation from 
the point of interaction $z$ to the final destination $y$. 

The propagation of the quark-antiquark pair in the shock-wave 
background is described by the product of two propagators of
Eq. (\ref{fla6}) type which contain two Wilson-line factors 
$U(z)U^{\dagger}(z')$ 
where $z'$ is the point where the antiquark crosses the shock wave. If we
substitute this quark-antiquark propagator in the original expression
for the amplitude (\ref{fla5}) we obtain\cite{ing}:
\begin{eqnarray}
&\int \! d^{4}x  d^{4}z e^{ip_A\cdot x +iq\cdot z} 
   \langle T\{j(x+z)j(z)\}\rangle_{A}
   \simeq\int 
   \frac {d^2p_{\perp}}{4\pi^2}
   I(p_{\perp},q_{\perp}) 
   {\mathop{\rm Tr}}\{ U(p_{\perp})U^{\dagger}(q_{\perp}-p_{\perp}) \} 
\nonumber
\end{eqnarray}
where $U(p_{\perp})$ is the Fourier transform of $U(x_{\perp})$ and 
$I(p_{\perp},q_{\perp})$ is the so-called ``impact factor" which is a 
function of $p_{\perp}^2,p_{\perp}\!\cdot\! q_{\perp}$, and photon 
virtuality \cite{mes},\cite{ing}. 
Thus, we have reproduced the leading term in the 
expansion (\ref{fla1}). (To recognize it, note that
$U(x_{\perp})U^{\dagger}(y_{\perp})=
P\exp\left\{-ig\int^x_y dz_i U_i(z_{\perp})\right\}$
where the precise form of the path between points
 $x_{\perp}$ and $y_{\perp}$ does not 
matter since this is actually a formula for the 
gauge link in a pure gauge field $U_i(z_{\perp})$).

 Note that formally we have obtained the operators $U$ ordered along the 
 light-like
lines. Matrix elements of such operators contain divergent 
longitudinal integrations which reflect the fact that light-like gauge factor 
corresponds to a quark moving with speed of light (i.e., with infinite 
energy). As demonstrated in \cite{ing}, we may regularize this 
divergence by changing the slope of the
supporting line: if we wish the longitudinal integration stop at
$\eta=\eta_0$, we should order our gauge factors $U$ along a line parallel
to $n=\sigma p_1+ \tilde{\sigma}p_2$.
Then  the coefficient 
functions in front of Wilson-line operators will contain logarithms 
$\sim g^2\ln 1/\sigma$. For example, there are corrections of such type 
to the impact factor $I(p,q)$ and if we sum them, the impact factor 
will be replaced by 
$\sum \left(g^2\ln 1/\sigma\right)^n{\cal K}^n I(p,q)$ where ${\cal K}$ is 
the BFKL kernel.  

\section*{Factorization formula for high-energy scattering}
In order to understand how this expansion can be generated by the factorization
formula of Eq. (\ref{fla3}) type we have to rederive the 
operator expansion in axial gauge $A_{\bullet}=0$ with an additional condition 
$\left.A_{*}\right|_{x_*=-\infty}=0$ (the existence of such a gauge was 
illustrated in\cite{wlup} by an explicit construction). It is important to 
note that with
with power accuracy (up to corrections $\sim \sigma$) our gauge condition may
be replaced by 
$e^{\mu}A_{\mu}=0$. In this gauge the coefficient
functions are given by Feynman diagrams in the external field
\begin{equation}
B_i(x)=U_i(x_{\perp})\Theta(x_*),~~~~~~~~~~~~~~~ B_{\bullet}=B_{*}=0
\label{fla10}
\end{equation} 
which is a gauge rotation of our shock wave (it is easy to see that the 
only nonzero component of the field strength tensor 
$F_{\bullet i}(x)=U_i(x_{\perp})\delta(x_*)$ corresponds to shock wave). 
The Green functions in external field (\ref{fla10}) can be obtained
from a generating functional with a source responsible for this external field. 
Normally, the source for given external field $\bar{{\cal A}}_{\mu}$ is just 
$J_{\nu}=\bar{D}^{\mu} \bar{F}_{\mu\nu}$ so in our case the only non-vanishing 
contribution  is $J_{*}(B)=\bar{D}^i\bar{F}_{i*}$. However, 
we have a problem because the field
which we try to create by this source does not decrease at infinity. To 
illustrate the problem, suppose that we use another light-like gauge
${\cal A}_{*}=0$ for a calculation of the propagators in the external field 
(\ref{fla10}). In this case, the only would-be nonzero contribution
to the source term in the functional integral 
$\bar{D}^i\bar{F}_{i_{\bullet}}{\cal A}_{*}$ vanishes,
 and it looks like 
we do not need a source at all to generate the field $B_{\mu}$!
(This is of course wrong since $B_{\mu}$ is not the classical solution).
What it really means is that the source in this case lies entirely at the 
infinity. Indeed, when we are trying to make an external field 
$\bar{{\cal A}}$ 
in the
functional integral by the source $J_{\mu}$ we need to make a shift
${\cal A}_{\mu}\rightarrow {\cal A}_{\mu}+\bar{{\cal A}}_{\mu}$ 
in the functional integral
\begin{eqnarray}
&\int{\cal D}{\cal A} \exp\left\{iS({\cal A})-i\!\int\! d^4x 
J^a_{\mu}(x){\cal A}^{a\mu}(x)\right\}
\label{fla11}
\end{eqnarray}
after which the linear term $\bar{D}^{\mu} \bar{F}_{\mu\nu}{\cal A}^{\nu}$ 
cancels
with our source term $J_{\mu}{\cal A}^{\mu}$ and the terms quadratic in 
${\cal A}$ 
make the Green functions in the external field $\bar {\cal A}$.
(Note that the classical action $S(\bar{{\cal A}})$ for our external 
field $\bar{{\cal A}}=B$ (\ref{fla10}) vanishes). 
However, in order to reduce the linear
term $\int d^4x\bar{F}^{\mu\nu}\bar{D}_{\mu}{\cal A}_{\nu}$ in the functional 
integral to the form 
$\int d^4x\bar{D}^{\mu} \bar{F}_{\mu\nu}{\cal A}^{\nu}(x)$ we need to make an 
integration by parts, and if the external field does not decrease 
there will be 
additional surface terms at infinity. In our case we are trying to make the
external field $\bar{{\cal A}}=B$ so the linear term which need to be 
canceled by the source is
\begin{eqnarray}
&{2\over s}\int\! dx_{\bullet}dx_{*}d^2x_{\perp} \bar{F}_{i\bullet}
\bar{D}_{*}{\cal A}^{i}=
\left.\int\! dx_{*}d^2x_{\perp} \bar{F}_{i\bullet}
{\cal A}^{i}\right|^{x_{\bullet}=\infty}_{x_{\bullet}=-\infty}
\label{fla12}
\end{eqnarray}
It comes entirely from the boundaries of integration. If we
recall that in our case 
$\bar{F}_{\bullet i}(x)=U_i(x_{\perp})\delta(x_*)$ we can finally rewrite 
the linear term as
\begin{eqnarray}
&\int\! d^2x_{\perp} U_i(x_{\perp})
\{{\cal A}^{i}(-\infty p_2+x_{\perp})-{\cal A}^{i}(\infty p_2+x_{\perp})\}
\label{fla13}
\end{eqnarray}
The source term which we must add to the exponent in the functional 
integral to cancel the linear term after the shift is given by Eq. (\ref{fla13})
with the minus sign. Thus, Feynman diagrams in the external
field (\ref{fla10}) in the light-like gauge ${\cal A}_{*}=0$ are generated
 by the functional integral
\begin{equation}
\!\int\!{\cal D}{\cal A} \exp\Big\{iS({\cal A})+
i\!\int\! d^2x_{\perp} 
U^{ai}(x_{\perp})[{\cal A}^a_i(\infty p_2+x_{\perp})
-{\cal A}^{ai}(-\infty p_2+x_{\perp})]\Big\}
\label{fla14}
\end{equation}
In an arbitrary gauge the source term in the exponent in Eq. (\ref{fla14}) 
can be rewritten in the form
\begin{eqnarray}
&2i\int d^2x_{\perp}{\mathop{\rm Tr}} \{U^i(x_{\perp})\int^{\infty}_{-\infty} 
dv[-\infty p_2+x_{\perp},vp_2+x_{\perp}]\nonumber\\
&
F_{*i}(vp_2+x_{\perp})[vp_2+x_{\perp},-\infty p_2+x_{\perp}]\}
\label{fla15}
\end{eqnarray}
Thus, we have found the generating functional for our Feynman diagrams in the 
external field (\ref{fla11}). However, it is easy to see (by inspection of the 
first rung of BFKL ladder diagram) that the longitudinal integrals over 
$\alpha$
in these diagrams will be unrestricted from below while we need the restriction
$\alpha>\sigma$. Fortunately, we already faced that problem on the other 
side -- 
in matrix elements of operators $U$ and we have solved it by changing 
the slope of the supporting  line. Similarly to the case of matrix elements, 
it can be demonstrated that if we want the logarithmical integrations 
over large $\alpha$ to stop at $\alpha=\sigma$, we need to order the gauge 
factors in 
Eq.(\ref{fla15}) along the same vector $n=\sigma p_1+\tilde{\sigma} p_2$, 
cf. Eq. (\ref{fla2}). Therefore,
the final form of the generating functional for the Feynman diagrams (with
 $\alpha>\sigma$ cutoff) in the external field  (\ref{fla11}) is
\begin{eqnarray}
\int\!{\cal D}{\cal A} {\cal D}\Psi
{\rm exp}\left\{iS({\cal A},\Psi)+
i\int d^2x_{\perp} U^{ai}(x_{\perp})V^a_i(x_{\perp})\right\}
\label{fla16}
\end{eqnarray}
where
\begin{eqnarray}
\lefteqn{V_i(x_{\perp})=}\label{fla17}\\
&\int^{\infty}_{-\infty} dv[-\infty
n+x_{\perp},vn+x_{\perp}] n^{\mu}F_{\mu i}(vn+x_{\perp})
[vn+x_{\perp},-\infty n+x_{\perp}]
\nonumber
\end{eqnarray}
and $V^a_i\equiv {\rm Tr}(\lambda^aV_i)$ as usual. For completeness, 
we have added 
integration over quark fields so $S({\cal A},\Psi)$ is the full QCD action.
 
 Now we can assemble the different parts of the factorization 
formula (\ref{fla4}). We have written down the generating functional integral 
for the diagrams with $\alpha>\sigma$ in the external fields with 
$\alpha<\sigma$ and what remains now is to write down the integral over 
these ``external'' fields. 
Since this
integral is completely independent of (\ref{fla16}) we will use a different
notation ${\cal B}$ and $\chi$ for the $\alpha<\sigma$ fields. We have: 
\begin{eqnarray}
\lefteqn{\int\!\! {\cal D}A{\cal D}\bar{\Psi}{\cal D}\Psi e^{iS(A,\Psi)} 
j(p_{A})j(p'_{A})j(p_{B})j(p'_{B})=}
\label{fla18}\\ 
&\int\!\! {\cal D}{\cal A}{\cal D}\bar{\psi}{\cal D}
\psi e^{iS({\cal A},\psi)} j(p_{A})j(p'_{A})
\int\!\! {\cal D}{\cal B}{\cal D}\bar{\chi}{\cal D}\chi \nonumber\\
&
j(p_{B})j(p'_{B})
e^{iS({\cal B},\chi)} \exp\Big\{i\!\int\! d^2x_{\perp} 
U^{ai}(x_{\perp})V^{a}_i(x_{\perp})\Big\}\nonumber
\end{eqnarray}
The operator $U_i$ in an arbitrary gauge is
given by the same formula (\ref{fla17}) as operator $V_i$
with the only difference that the gauge links and $F_{{\bullet} i}$ 
are constructed from the fields 
${\cal B}_{\mu}$. This is our main result (\ref{fla4}) 
in the functional integral representation.

The functional integrals over ${\cal A}$ fields give logarithms of the
type $g^2\ln{1/\sigma}$ while the integrals over slow ${\cal B}$ fields give
powers of $g^2\ln (\sigma s/m^2)$. With logarithmic accuracy, they add up to
$g^2\ln s/m^2$. However, there will be
additional terms $\sim g^2$ due to mismatch coming from the region 
of integration near the dividing point $\alpha\sim\sigma$ where the
details of the cutoff in the matrix elements of the operators $U$ and $V$ 
become important. Therefore, one should expect the corrections of order of 
$g^2$ to the effective action $\int dx_{\perp} U^iV_i$. Still,
the fact that the fast quark moves along the straight line has nothing
to do with perturbation theory (cf. ref. \cite{dosch}); therefore it is 
natural to expect the
non-perturbative generalization of the factorization formula (\ref{fla18}) 
constructed from the same Wilson-line operators $U_i$ and $V_i$
(probably with some kind of non-local interactions between them).

\section{Effective action for high-energy scattering}
The factorization formula gives us a starting point for a new approach
to the analysis of the high-energy effective action. 
Consider another rapidity $\eta'_0$ in the region between $\eta_0$ and 
$\ln m^2/s$. If we use the factorization formula 
(\ref{fla18}) once more, this time dividing between the rapidities 
greater and smaller than $\eta'_0$, we get the expression 
for the amplitude (\ref{fla5}) in the form:
\begin{eqnarray}
iA(s,t)&=&\int\!\! {\cal D}Ae^{iS(A)} 
j(p_{A})j(p'_{A})j(p_{B})j(p'_{B})
\label{fla19}\\ 
&=&\int\!\! {\cal D}{\cal A} e^{iS({\cal A})} j(p_{A})j(p'_{A})
\int\!\! {\cal D}{\cal B}
e^{iS({\cal B})}j(p_{B})j(p'_{B}) \nonumber\\
&~&\int\! {\cal D}{\cal C}e^{iS({\cal C})}
e^{i\!\int\! d^2x_{\perp} 
V^{ai}(x_{\perp})U^a_i(x_{\perp})+i\!\int\! d^2x_{\perp} 
W^{ai}(x_{\perp})Y^a_i(x_{\perp})}
\nonumber
\end{eqnarray}
(For brevity, we do not display the quark fields). 
In this formula operators $V_i$ (made from ${\cal A}$) fields 
are given by Eq. (\ref{fla17}), the operators $U_i$ are also given by 
Eq. (\ref{fla17}) but constructed from 
${\cal C}$ fields, and the operators $W_i$ (made from ${\cal C}$ fields) and $Y_i$ 
(made from ${\cal B}$ fields) are aligned along the 
direction $n'=\sigma'p_1+\tilde{\sigma}'p_2$ 
corresponding to the rapidity $\eta'$ (as usual, 
$\ln \sigma'/\tilde{\sigma}'=\eta'$ 
where
$\tilde{\sigma}'=m^2/s\sigma'$):
\begin{eqnarray}
&U_i(x_{\perp})=\int^{\infty}_{-\infty} dv[-\infty n+x_{\perp},vn+x_{\perp}]
n^{\mu}F_{\mu i}(vn+x_{\perp})
[vn+x_{\perp},-\infty n+x_{\perp}]_{\cal C}\nonumber\\
&W_i(x_{\perp})=\int^{\infty}_{-\infty} dv[-\infty n'+x_{\perp},vn'+x_{\perp}]
n^{'\mu}F_{\mu i}(vn'+x_{\perp})
[vn'+x_{\perp},-\infty n'+x_{\perp}]_{\cal C}\nonumber\\
&Y_i(x_{\perp})=\int^{\infty}_{-\infty} dv[-\infty n'+x_{\perp},vn'+x_{\perp}]
n^{'\mu}F_{\mu i}(vn'+x_{\perp})
[vn'+x_{\perp},-\infty n'+x_{\perp}]_{\cal B}
\nonumber
\end{eqnarray}
Thus, we have factorized the functional integral 
over ``old'' ${\cal B}$ fields 
into the product of two integrals over ${\cal C}$ and ``new" ${\cal B}$
fields.

Now, let us integrate over the ${\cal C}$ fields  
and write down the result in terms of an effective action. 
Formally, one obtains:
\begin{equation}
iA(s,t)=
\int\!\! {\cal D}{\cal A} e^{iS({\cal A})} j(p_{A})j(p'_{A})
\int\!\! {\cal D}{\cal B}
e^{iS({\cal B})}j(p_{B})j(p'_{B})e^{iS_{\rm eff}(V_i,Y_i;{\sigma\over\sigma'})} 
\label{fla21}
\end{equation}
where $S_{\rm eff}$ for the rapidity interval between $\eta$ and 
$\eta'$ is defined as 
\begin{eqnarray}
&e^{iS_{\rm eff}(V_i,Y_i;{\sigma\over\sigma'})}=\int\! {\cal D}{\cal C}e^{iS({\cal C})}
e^{i\!\int\! d^2x_{\perp} 
V^{ai}(x_{\perp})U^a_i(x_{\perp})+i\!\int\! d^2x_{\perp} 
W^{ai}(x_{\perp})Y^a_i(x_{\perp})}\label{fla22}
\end{eqnarray}
This formula gives a rigorous definition for the effective action for a 
given interval in rapidity 
(cf. ref. \cite{lobzor}).
Next step would be to perform explicitly the integrations over the 
longitudinal momenta in the r.h.s. 
of Eq. (\ref{fla22}) and obtain the answer
for the integration over
our rapidity region (from $\eta$ to $\eta'$) in terms of two-dimensional 
theory in the transverse coordinate space which hopefully would give us 
the unitarization of the BFKL pomeron. At present, it is not 
known how to do this. One can obtain, however, a first few terms in the 
expansion of effective action in powers of $V_i$ and $Y_i$. The easiest way 
to do this is to expand gauge factors $U_i$ and $W_i$ in r.h.s. of Eq. 
(\ref{fla22}) in powers of ${\cal C}$ fields and calculate the relevant 
perturbative diagrams (see Fig.2)..
\begin{figure}[t]
\hspace{1cm}
\psfig{figure=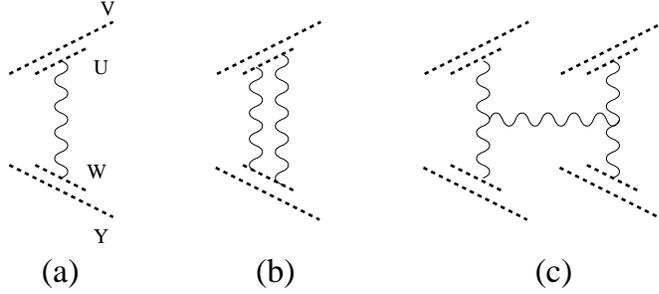,height=1.5in}
\caption{Perturbative expansion of effective action.\label{fig:2}}
\end{figure}
For illustration, let us present a couple of first terms
in the effective  action \cite{Verlinde},\cite{many}:
\begin{eqnarray}
\lefteqn{S_{\rm eff} = \int d^2x 
V^{ai}(x)Y^a_i(x)-}\label{fla23}\\
&{g^2\over 64\pi^3}\ln{\sigma\over\sigma'}
\Big(N_c\int d^2x d^2y V^a_{i,i}(x)\ln^2(x-y)^2Y^a_{j,j}(y)+
{f_{abc}f_{mnc}\over 4\pi^2}\!\int\! d^2x d^2y d^2x' d^2y'\nonumber\\
&
V^a_{i,i}(x)V^m_{j,j}(y)Y^b_{k,k}(x')Y^n_{l,l}(y')
\ln{(x-z)^2\over(x-x')^2}\ln{(y-z)^2\over(y-y')^2}
\left({\partial\over\partial z_i}\right)^2\ln{(x'-z)^2\over(x-x')^2}
\ln{(y'-z)^2\over(y-y')^2}\Big) +...
\nonumber
\end{eqnarray}
where we we use the notation 
$V^a_{i,j}(x)\equiv {\partial\over \partial x_j}V^a_i(x)$ etc. 
The first term (see Fig. 2a) looks like the corresponding term in the 
factorization 
formula (\ref{fla18}) -- only the directions of the supporting lines are 
now strongly different.  
The second
term shown in Fig. 2b is the first-order expression for the 
reggeization of the gluon\cite{fkl}
and the third term (see Fig. 2c) is the two-reggeon  
Lipatov's Hamiltonian\cite{l} responsible for BFKL logarithms. 

\section{Effective action and collision of two shock waves} 

The functional integral (\ref{fla22})
which defines the effective action is the usual QCD functional integral 
with two sources corresponding to the two colliding shock waves. 
Instead of calculation of perturbative diagrams 
(as it was done in previous section) one can use the
semiclassical approach. This approach
is relevant when the coupling constant is relatively small but the 
characteristic fields are large (in other words, when $g^2\ll 1$ 
but $gV_i\sim gY_i \sim 1$). In this case one can 
calculate the functional integral (\ref{fla22}) by expansion around the new
stationary point corresponding to the classical
wave created by the collision of the shock waves.

With leading log accuracy, we can replace the vector $n$ by $p_1$ and the
vector $n'$ by $p_2$. 
The classical equations for the wave created by the collision are:
\begin{eqnarray}
\lefteqn{D^{\mu}F_{\mu i}=0}\label{fla24}\\
&D^{\mu}F_{\bullet\mu}=
\delta({2\over s}x_{\ast})[{2\over s}x_{\bullet}p_2+x_{\perp}, 
-\infty p_2+x_{\perp}]\nabla_i Y^i(x_{\perp})
[-\infty p_2+x_{\perp},{2\over s}x_{\bullet}p_2+x_{\perp}]\nonumber\\
&D^{\mu}F_{\ast\mu}=
\delta({2\over s}x_{\bullet})[{2\over s}x_{\ast}p_1+x_{\perp}, 
-\infty p_1+x_{\perp}]\nabla_i V^i(x_{\perp})
[-\infty p_1+x_{\perp},{2\over s}x_{\ast}p_2+x_{\perp}]\nonumber
\end{eqnarray}
where
\begin{eqnarray}
\nabla_iY^i&\equiv&\partial_iY^i\nonumber\\
&-&i\Big[\int_{-\infty}^{\infty}dv 
[-\infty p_2+x_{\perp},vp_2+x_{\perp}]F_{\ast i}(vp_2+x_{\perp})
[vp_2+x_{\perp},-\infty p_2+x_{\perp}],Y^i\Big]\nonumber\\
\nabla_iV^i&\equiv&\partial_iV^i\nonumber\\
&-&i\Big[\int_{-\infty}^{\infty}dv 
[-\infty p_1+x_{\perp},vp_1+x_{\perp}]
F_{\bullet i}(vp_1+x_{\perp})
[vp_1+x_{\perp},-\infty p_1+x_{\perp}],V^i\Big]
\nonumber
\end{eqnarray}
The r.h.s of the Eq. (\ref{fla24}) is the first-order 
variational derivative of the source terms $\int\! d^2x_{\perp} 
V^{ai}(x_{\perp})U^a_i(x_{\perp})$ and $\int\! d^2x_{\perp} 
W^{ai}(x_{\perp})Y^a_i(x_{\perp})$ with respect to the gauge field.
Also, as explained in Sect. 3, because our fields do not decrease at
infinity there may be extra surface linear terms (cf. Eq. (\ref{fla12})). 
The requirement
of absence of such terms gives four additional equations 
\begin{eqnarray}
\left.F_{\bullet i}\right|_{x_{\bullet}=-\infty}&=&
\delta(2x_{\ast}/s)Y_i(x_{\perp}),~~~~~~~
\left.F_{\ast i}\right|_{x_{\ast}=-\infty}= \delta(2x_{\bullet}/s)
V_i(x_{\perp}),\label{fla26}\\ 
\left.F_{\bullet i}\right|_{x_{\bullet}=\infty}&=&\delta({2\over s}x_{\ast})
[\infty p_2+x_{\perp}, -\infty p_2+x_{\perp}] Y_i(x_{\perp}) [-\infty
p_2+x_{\perp}, \infty p_2+x_{\perp}]\nonumber\\ 
\left.F_{\ast i}\right|_{x_{\ast}=\infty}&=&\delta({2\over s}x_{\bullet})
[\infty p_1+x_{\perp}, -\infty p_1+x_{\perp}] V_i(x_{\perp})
[-\infty p_1+x_{\perp}, \infty p_1+x_{\perp}]
\nonumber
\end{eqnarray}
The two sets (\ref{fla24}) and (\ref{fla26})
define the classical field created by the collision of two shock waves.

Unfortunately, it is not clear how to solve these equations. One can 
start with the trial field
which is a simple superposition of the two shock waves (\ref{fla10}) 
\begin{equation}
A^{(0)}_{\ast}=A^{(0)}_{\bullet}=0,~~~~
A^{(0)}_i=\Theta(x_{\bullet})V_i+\Theta(x_{\ast})Y_i
\label{fla27}
\end{equation}
and improve
it by taking into account the interaction between the shock waves order by
order (cf. ref. \cite{many}). The parameter of this expansion is the 
commutator $g^2[Y_i,V_k]$. Moreover,  it can be demonstrated that 
each extra commutator brings a factor
$\ln{\sigma\over\sigma'}$ and therefore this approach is a sort of
leading logarithmic approximation.  In the lowest nontrivial order one gets:
\begin{eqnarray} 
A^{(1)}_i&=&-{g\over 4\pi^2}\int
dz_{\perp}([Y_i(z_{\perp}),V_k(z_{\perp})]-i\leftrightarrow k)
{(x-z)^k\over(x-z)_{\perp}^2}\ln\left(1-{(x-z)_{\perp}^2\over
x_{\parallel}^2+i\epsilon}\right) \nonumber\\ 
A^{(1)}_{\bullet}&=&{gs\over 16\pi^2}\int
dz_{\perp} {1\over
x_{\ast}-i\epsilon}\ln(-x_{\parallel}^2+(x-z)_{\perp}^2+i\epsilon)
[Y_k(z_{\perp}),V^k(z_{\perp})] \nonumber\\ 
A^{(1)}_{\ast}&=&-{gs\over 16\pi^2}\int dz_{\perp}
{1\over x_{\bullet}-i\epsilon}\ln(-x_{\parallel}^2+(x-z)_{\perp}^2+i\epsilon)
[Y_k(z_{\perp}),V^k(z_{\perp})] \label{fla28}
\end{eqnarray}
where $x_{\parallel}^2\equiv{4\over s}x_{\ast}x_{\bullet}$ is a 
longitudinal part of $x^2$. These fields are
obtained in the background-Feynman gauge. The corresponding expressions for
field strength have the form:
\begin{eqnarray}
F^{(1)}_{\bullet\ast}&=&{gs\over 4\pi^2}
\int dz_{\perp}{1\over -x_{\parallel}^2+(x-z)_{\perp}^2+i\epsilon}[Y_k,V^k]
\label{fla29}\\
F^{(1)}_{ik}&=&{g\over 2\pi^2}
\int dz_{\perp}{1\over -x_{\parallel}^2+(x-z)_{\perp}^2+i\epsilon}
([Y_i,V_k]-[Y_k,V_i])\nonumber\\
F^{(1)}_{\bullet i}&=&
{gs\over 8\pi^2}
\!\int\! dz_{\perp}{(x-z)^k\over -x_{\parallel}^2+(x-z)_{\perp}^2+i\epsilon}
\left({g_{ik}[Y_j,V^j]\over x_{\ast}-i\epsilon}+
{[Y_i,V_k]-[Y_k,V_i]\over
x_{\ast}+i\epsilon}\right) \nonumber\\
F^{(1)}_{\ast i}&=&-
{gs\over 8\pi^2}
\!\int\! dz_{\perp}{(x-z)^k\over -x_{\parallel}^2+(x-z)_{\perp}^2+i\epsilon}
\left({g_{ik}[Y_j,V^j]\over x_{\bullet}-i\epsilon}-
{[Y_i,V_k]-[Y_k,V_i]\over
x_{\bullet}+i\epsilon}\right)\nonumber
\end{eqnarray}

Let us now find the effective action. In the trivial order the only non-zero 
field strength components are 
$F^{(0)}_{\bullet i}=\delta({2\over s}x_{\ast})Y_i(x_{\perp})$ and
$F^{(0)}_{\ast i}=\delta({2\over s}x_{\bullet})V_i(x_{\perp})$ 
so we get the familiar  expression 
$S^{(0)}=\int d^2x_{\perp}V^{ai}Y^{a}_i$. In the next order one has
\begin{eqnarray}
&&S^{(1)}=\int d^4x\left(-{2\over s}F_{\ast}^{(1)ai}F^{(1)a}_{\bullet i}-
{1\over 4}F^{(1)a}_{ik}F^{(1)aik}+{2\over s^2}F^{(1)a}_{\ast\bullet}
F^{(1)a}_{\ast\bullet}\right)+2\!\int\! d^2x_{\perp} du\nonumber\\
&&
\Big({\rm Tr}V^{i}
\left([-\infty p_2+x_{\perp},up_2+x_{\perp}] 
F_{\bullet i}(up_1+x_{\perp})
[up_2+x_{\perp},-\infty p_2+x_{\perp}]\right)^{(1)}+\nonumber\\
&&
{\rm Tr}Y^{i} 
\left([-\infty p_1+x_{\perp},up_1+x_{\perp}]
F_{\ast i}(up_1+x_{\perp})[up_1+x_{\perp},-\infty
p_1+x_{\perp}]\right)^{(1)}\Big) \label{fla30}
\end{eqnarray} 
We have seen above that the effective action
contains $\ln{\sigma\over\sigma'}$ (see Eq. (\ref{fla23})). 
With logarithmic accuracy the r.h.s of Eq. (\ref{fla30}) reduces to
\begin{eqnarray}
S^{(1)}&=&-{2\over s}\int d^4x F^{(1)ai}_{\ast}(x)F_{\bullet i}^{(1)a}(x)
\nonumber\\
&+&\int d^2x_{\perp}2{\rm Tr}[Y^{i},V_{i}]
\left([x_{\perp},-\infty p_2+x_{\perp}]^{(1)}-
[x_{\perp},-\infty p_1+x_{\perp}]^{(1)}\right)
\label{fla31}
\end{eqnarray}
The first term contains the integral over
$d^4x={2\over s}dx_{\bullet}dx_{\ast}d^2x_{\perp}$. In order to separate the 
longitudinal divergencies 
from the infrared divergencies in the transverse space we will
work in the $d=2+2\epsilon$ transverse dimensions. 
It is convenient to perform at first the 
integral over $x_{\ast}$ which is determined by a residue in the
point $x_{\ast}=0$. The integration over remaining 
light-cone variable 
$x_{\bullet}$ factorizes then in the form 
$\int_{0}^{\infty}dx_{\bullet}/x_{\bullet}$ or
$\int_{-\infty}^{0}dx_{\bullet}/x_{\bullet}$.
This integral reflects our usual longitudinal logarithmic divergencies
which arise from the replacement of vectors $n$ and $n'$ in (\ref{fla22}) 
by the light-like vectors $p_1$ and $p_2$. 
In the momentum space this logatithmical divergency has the form 
$\int d\alpha/\alpha$. 
It is clear that when $\alpha$ is close to $\sigma$ (or $\sigma'$) we 
can no longer approximate $n$ by $p_1$ (or $n'$ by $p_2$). Therefore, 
in the leading log accuracy this divergency should be replaced by $
\ln{\sigma\over\sigma'}$:
\begin{eqnarray}
\int_{0}^{\infty}dx_{\bullet}{1\over x_{\bullet}}=\int_{0}^{\infty}d\alpha {1\over \alpha}\rightarrow 
\int_{\sigma}^{\sigma'}d\alpha {1\over \alpha}~=~\ln{\sigma\over\sigma'}
\label{fla33}
\end{eqnarray}
The (first-order) gauge links in the second term in r.h.s. 
of Eq. (\ref{fla31}) have the logarithmic divergence of the same origin:
\begin{eqnarray}
&&[x_{\perp},-\infty p_1+x_{\perp}]^{(1)}=-{i\over 8\pi^2}
\int_{-\infty}^{0}dx_{\ast}{1\over x_{\ast}}\int d^2x_{\perp}
{\Gamma(\epsilon)\over(x-z)_{\perp}^{2\epsilon}}
[Y_k(z_{\perp}),V^k(z_{\perp})]\nonumber\\
&&[x_{\perp},-\infty p_2+x_{\perp}]^{(1)}={i\over 8\pi^2}
\int_{-\infty}^{0}dx_{\bullet}{1\over x_{\bullet}}\int d^2x_{\perp}
{\Gamma(\epsilon)\over(x-z)_{\perp}^{2\epsilon}}
[Y_k(z_{\perp}),V^k(z_{\perp})]
\nonumber
\end{eqnarray}
which also should be replaced by $\ln{\sigma\over\sigma'}$.
Performing the remaining integration over $x_{\perp}$ in the first term
in r.h.s. of Eq. (\ref{fla31}) we obtain the 
the first-order classical action in the form:
\begin{eqnarray}
S^{(1)}&=&{ig^2\over 16\pi^2}\ln{\sigma\over\sigma'}
\int d^2x_{\perp} d^2y_{\perp} L^a_{ik}(x_{\perp})
{\Gamma(\epsilon)\over(x-y)_{\perp}^{2\epsilon}}
L^{aik}(y_{\perp})
\label{fla34}\\
{\rm where}~~~~~~~~~~~~&~&\nonumber\\
L^a_{ik}&=&f^{abc}(Y^a_jV^{bj}g_{ik}+Y^a_iV^b_k-Y^a_kV^b_i)
\label{fla35}
\end{eqnarray}
At $d=2$ we have an infrared pole in $S^{(1)}$ which must be cancelled by the 
corresponding divergency in the trajectory of the reggeized gluon. 
The gluon reggeization is not a classical effect in our approach - rather, 
it is a 
first quantum correction to our classical field (\ref{fla28}). 
The relevant term can be obtained using the
evolution equations for operators $U$ from Ref. \cite{ing}. 
One gets:
\begin{eqnarray}
\lefteqn{S_{\rm r}=}\label{fla36}\\
&-{g^2N_c\over 16\pi^3}\ln{\sigma\over\sigma'}
\int d^2x_{\perp} d^2y_{\perp} (V_i^a(x_{\perp})-V_i^a(y_{\perp}))
{\Gamma^2(1+\epsilon)\over((x-y)_{\perp}^2)^{(1+2\epsilon)}}
(Y^{ai}(x_{\perp})-Y^{ai}(y_{\perp}))\nonumber
\end{eqnarray}
This expression coincides with the second term
in r.h.s. of Eq. (\ref{fla23})) up to the terms proportional to 
higher commutators which we neglect here.

Thus, the first-order expression for the effective action is
the sum of $S^{(0)}$, $S^{(1)}$, and $S_{\rm r}$:
\begin{eqnarray}
\lefteqn{S_{\rm eff}=
\int d^2x_{\perp}V^{ai}(x_{\perp})Y^{a}_i(x_{\perp})+}\label{fla37}\\
&&{g^2\over 16\pi^2}\ln{\sigma\over\sigma'}\left\{i\int d^2x_{\perp}
d^2y_{\perp}
L^a_{ik}(x_{\perp}){\Gamma(\epsilon)\over(x-z)_{\perp}^{2\epsilon}}
L^{aik}(y_{\perp}) -\right.\nonumber\\ &&\left.{N_c\over \pi}
\int d^2x_{\perp} d^2y_{\perp}(V_i^a(x_{\perp})-V_i^a(y_{\perp}))
{\Gamma^2(1+\epsilon)\over((x-y)_{\perp}^2)^{(1+2\epsilon)}}
(Y^{ai}(x_{\perp})-Y^{ai}(y_{\perp}))\right\} \nonumber
\end{eqnarray}
which coincide with (\ref{fla23}) up to the higher commutators. 
As usual, in the case of scattering of white objects the logarithmic
infrared divergence $\sim {1\over\epsilon}$ cancels. 
For example, for the case of one-pomeron exchange the relevant term
in the expansion of effective action is
\begin{eqnarray}
&-{g^2\over 16\pi^2}\ln{\sigma\over\sigma'}\int d^2x_{\perp} d^2y_{\perp} 
f^{dam}(V_j^aY^{mj}g_{ik}+V^a_iY^m_k-V^a_kY^m_i)(x_{\perp})\nonumber\\
&
{\Gamma(\epsilon)\over(x-y)_{\perp}^{2\epsilon}}
f^{dbn}(V_l^bY^{nl}g^{ik}+V^{bi}Y^{mk}-V^{bk}Y^{mi})(y_{\perp})+\nonumber\\
&
{g^2N_c\over 16\pi^3}\ln{\sigma\over\sigma'}
\int d^2x_{\perp}V^a_i(x_{\perp})Y^{ai}(x_{\perp})\int d^2y_{\perp} 
d^2y'_{\perp} (V_j^b(y_{\perp})-V_j^b(y'_{\perp}))
\nonumber\\
&{\Gamma^2(1+\epsilon)\over((y-y')_{\perp}^2)^{(1+2\epsilon)}}
(Y^{bj}(y_{\perp})-Y^{bj}(y'_{\perp}))
\label{fla38}
\end{eqnarray}
It is easy to see that the terms $\sim {1\over\epsilon}$  cancel if we project onto 
colorless state in t-channel (that is, replace $V^{ai}V_j^b$ by 
${\delta_{ab}\over N_c^2-1}V^{ci}V_j^c$). It is worth noting that in the two-gluon
approximation the r.h.s. of the eq. (\ref{fla38}) gives the BFKL kernel.

In conclusion let us mention that this semiclassical approach is 
suited for the study of the heavy-ion collisions.  Indeed, for 
heavy-ion collisions the coupling constant 
may be relatively small due to high density 
(see \cite{larry2}). 
On the other hand, the fields produced by colliding ions are large so 
that the product $gA$ is not small -- 
which means that the Wilson-line gauge factors $V$ and $Y$ are of order of 1. 
In this case we need to know not only a couple of the first few terms in 
the expansion of the effective action, but the whole series.

\vskip0.5cm
\section*{Acknowledgments}

The author is grateful to L.N. Lipatov and A.V. Radyushkin for valuable 
discussions. 
This work was supported by the US Department of Energy under contract 
DE-AC05-84ER40150.

\end{document}